\documentclass[twocolumn, twocolappendix]{aastex701}

\def\ion[#1 #2]{#1\,{\sc #2}}
\newcommand{\Rs}{$ R_{\odot} $}

\begin{document}

\title{The 2023 Australian Total Solar Eclipse: Line Emission of Fe XIV, Fe X and Fe XI out to 6 solar radii}

\author[orcid=0000-0002-6396-8209, gname='Benjamin', sname='Boe']{Benjamin Boe}
\affiliation{Wentworth Institute of Technology, Boston, MA 02115, USA}
\email[show]{boeb@wit.edu}  

\author[orcid=0000-0003-4089-9316, gname='Shadia', sname='Habbal']{Shadia Habbal}
\affiliation{Institute for Astronomy, University of Hawaii, Honolulu, HI 96822, USA}
\email{habbal@hawaii.edu}  

\author[orcid=0000-0001-7312-2410, gname='Miloslav', sname='Druckm\"uller']{Miloslav Druckm\"uller}
\affiliation{Faculty of Mechanical Engineering, Brno University of Technology, Technicka 2, 616 69 Brno, Czech Republic}
\email{druckmuller@fme.vutbr.cz}  

\author[orcid=0000-0001-8150-913X, gname='Pavel', sname='\v Starha']{Pavel \v Starha}
\affiliation{Faculty of Mechanical Engineering, Brno University of Technology, Technicka 2, 616 69 Brno, Czech Republic}
\email{starha@fme.vutbr.cz}  

\author[orcid=0009-0003-5328-6743, gname='Mat\v ej', sname='\v Starha']{Mat\v ej \v Starha}
\affiliation{Faculty of Mechanical Engineering, Brno University of Technology, Technicka 2, 616 69 Brno, Czech Republic}
\email{starhamatej@gmail.com}  

\author[orcid=0000-0003-4598-3402, gname='Jana', sname='Hoderov\'a']{Jana Hoderov\'a}
\affiliation{Faculty of Mechanical Engineering, Brno University of Technology, Technicka 2, 616 69 Brno, Czech Republic}
\email{hoderova@fme.vutbr.cz}  

\author[orcid=0000-0002-8937-5620, gname='Sage', sname='Constantinou']{Sage Constantinou}
\affiliation{Institute for Astronomy, University of Hawaii, Honolulu, HI 96822, USA}
\email{sagecons@hawaii.edu}  

\author[orcid=0000-0003-2150-6935, gname='Eric', sname='Ayars']{Eric Ayars}
\affiliation{Department of Physics, California State University, Chico, CA 95929, USA}
\email{eayars@csuchico.edu}  
 
\author[gname='Daniell', sname='Casillas']{Daniell Casillas}
\affiliation{Department of Physics, California State University, Chico, CA 95929, USA}
\email{danixcasi@gmail.com}

\begin{abstract}
We present narrowband observations of the \ion[Fe xiv] (530.3 nm), \ion[Fe x] (637.4 nm), and \ion[Fe xi] (789.2 nm) coronal emission lines from the 2023 April 20 Total Solar Eclipse in Australia. We deployed pairs of telescopes for each emission line that were equipped with narrowband filters centered on, and several nanometers away from, the center wavelengths of the lines. The secondary continuum telescopes were used to measure and remove the combined continuum K- (electron) and F- (dust) corona, which dominate coronal emission at optical and infrared wavelengths. Significant emission was detected from all three lines from 1.03 solar radii (\Rs) continuously outward to at least 6 \Rs. The brightness of the lines and continuum are absolutely calibrated to the solar disk, and are validated by a comparison with LASCO-C2 observations made at the same time. Using these observations, we inferred the line emission ratios resolved throughout the middle-corona (defined as 1.5 to 6 \Rs) for the first time. These line ratios are a probe of the electron temperature, which have important implications for constraining models of coronal heating and the characterization of solar wind formation, yet these emission lines have scarcely been quantified beyond 3 \Rs \ in the corona. This study demonstrates the enduring potential of eclipse observations for coronal physics and suggests that future spacecraft missions could observe these lines farther out than has been attempted previously.
\end{abstract}

%\uat{Solar coronal holes}{1484}

\section{Introduction} 
\label{sec:Intro}

\begin{figure*}[ht!]
\centering
\includegraphics[width= 0.95\linewidth]{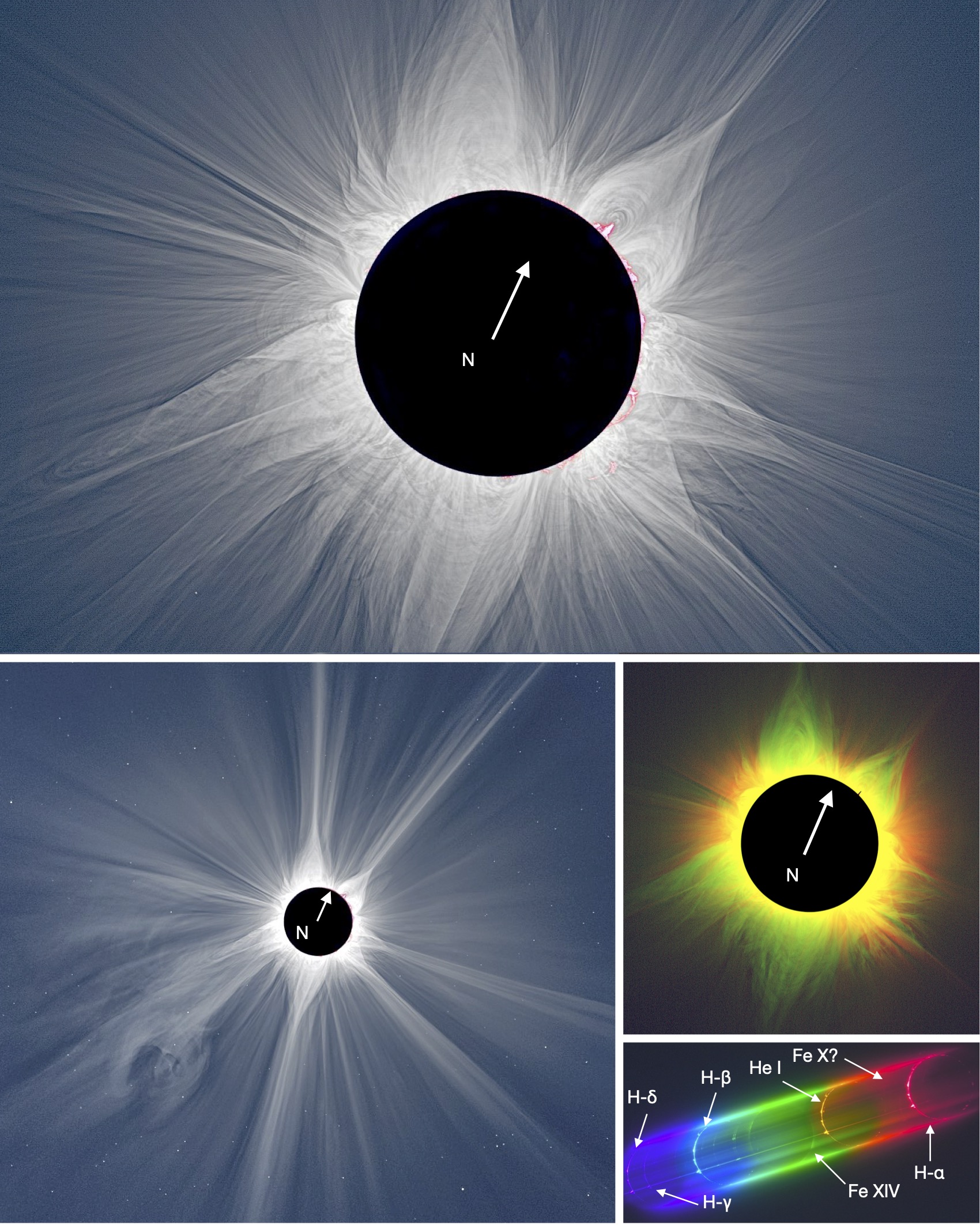}
\caption{White-light and narrowband images of the corona during the TSE from Exmouth, Australia, on 2023 April 20. The top image was taken with a 1000mm lens, while the bottom-left image shows a wider field-of-view from a 200mm lens (composited with the 1000mm image for the low corona). The middle-right panel shows a composite of \ion[Fe xiv] (green) and \ion[Fe x] (red) emission (see Section \ref{sec:Narrow}). The arrows in each image indicate the direction of solar North. The bottom right panel shows a slit-less flash spectrum of the corona with the strong Hydrogen, Helium, and Iron lines labeled.}
\label{fig:Whitelight}
\end{figure*}

Observations of ionic emission lines have long been used to characterize the physical properties of the solar corona. Ionic line emission was first observed in the corona during the 1869 Total Solar Eclipse (TSE) by \cite{Young1872}, using a slit-less spectrum (similar to the modern reproduction shown in the bottom right panel of Figure \ref{fig:Whitelight}). Originally thought to be a new element named ``Coronium", this ``Green" line at 530.3 nm, as well as the ``Red" coronal line at 637.4 nm, were subsequently identified as \ion[Fe xiv] and \ion[Fe x] respectively by \cite{Grotrian1939} and \cite{Edlen1943}. 

\par
For several decades, coronal line emission could only be seen during TSEs until the invention of the coronagraph by Lyot, who then identified several coronal lines \citep{Lyot1939}, including a near-infrared line at 789.2 nm of \ion[Fe xi]. Surveys of \ion[Fe xiv] (and often \ion[Fe x]) continued for decades throughout the 20th-century at Lomnick\'y \v{s}t\'{i}t, Sacramento Peak, and others (e.g., \citealt{Rybansky1994, Altrock2011, Oloketuyi2024}), albeit at a low coronal height of 1.15 solar radii (\Rs). Eclipse-based observations continued as well during this time, though the emission detected was typically below 1.3 \Rs \ \citep{Guhathakurta1992}, with occasional detections out to 1.7 \Rs \ \citep{Singh1982}.

\par

Based on these foundational findings, the Large-Angle Spectrometric COronagraph (LASCO) C1 coronagraph (LASCO-C1; \citealt{Brueckner1995}) onboard the Solar and Heliospheric Observatory (SOHO), included a Fabry-P\'erot interferometer to observe these visible-wavelength lines, which produced measurements of the line brightness \citep{Wang1997, Srivastava2000} and line-widths \citep{Mierla2008} of both \ion[Fe xiv] and \ion[Fe x] until its premature failure in 1998. These data were still limited in helioprojective height to less than about 2 \Rs, but demonstrated the strong potential of observing these lines from space. 

\par
Much of the focus on these visible wavelength lines diminished in the early 21st century, largely due to the introduction of observations from a wider spectrum of wavelengths. At somewhat shorter wavelengths in the ultraviolet (ranging from $\approx$ 47 to 135 nm), several coronal ionic lines as well as Lyman Alpha (Ly-$\alpha$, 121.6 nm) were spectroscopically observed by SOHO's Ultraviolet Coronagraph Spectrometer (UVCS; \citealt{Kohl1995}) consistently out to 3.5-5 \Rs \ \citep{Raymond1997, Ko1997, Suleiman1999, Akmal2001, Akinari2008}, with some detections in the tips of streamers out to as far as 8 \Rs \ \citep{Kohl1997}. 

\par
The new coronagraph Metis \citep{Antonucci2020} on Solar Orbiter has expanded these observations, imaging Ly-$\alpha$ emission out to 7 \Rs \ in streamer tips near solar-minimum \citep{Romoli2021}, and from CME material out to heights as much as 9 \Rs \ (which is claimed to translate to 23 \Rs \ after accounting for projection effects; \citealt{Russano2024}). 

\par

While Ly-$\alpha$ is the strongest resonantly excited line in the corona, and provides a useful method for probing the solar wind speed due to the ``Doppler-dimming" effect (e.g., \citealt{Hyder1970, Withbroe1982, Kohl1998, Giordano2025}), it cannot directly probe the electron temperature, unlike what is achievable with ionic line emission from heavier elements. Still, it may be possible to use polarized observations of Ly-$\alpha$ to measure the magnetic field strength via the Hanle effect, though likely only near active regions \citep{Raouafi2016, Supriya2021}.

\par

Now equipped with space-based telescopes, observations at Extreme Ultraviolet (EUV) and X-ray wavelengths have ushered in a new era of coronal astronomy. The abundant quantities of these data in the lower regions of the corona (often below 1.5 \Rs) have proven exceptionally useful for a wide range of applications (see \citealt{DelZanna2018} and ref. therein). In particular, these observations have helped inform coronal heating and solar wind acceleration models by constraining the electron temperature distributions in various structures (see \citealt{Viall2021}), and to investigate the initiation of coronal mass ejections (CMEs; see \citealt{Thompson2021}) and other complex dynamics near the Sun. 

\par
One drawback of emission at EUV and X-ray wavelengths in particular is that they fade much more rapidly with heliocentric distance than do lines at near-UV, visible, and infrared wavelengths. Low down in the corona, collisions between electrons and ions cause lines to be excited regardless of incoming radiation. Farther out in the corona (typically at about 1.5 to 2.5 \Rs;\ e.g., \citealt{Habbal2007, Seaton2021, Boe2022}), the rate of collisions drops dramatically due to the steep drop off in density with heliocentric distance, and so the emission generated by collisions reduces dramatically. While the brightness from collisional excitation depends on the density squared ($\propto n_e^2$; requires two particle interaction), the brightness due to radiative excitation depends linearly on density ($\propto n_e$). So, for wavelengths in the near-UV, visible, and infrared, the solar photosphere offers an abundant source of incident radiation to excite the lines, enabling them to be observed out to high helioprojective distances (e.g., \citealt{Habbal2010a, Habbal2011}). 

\par

There is some radiative excitation of the EUV lines due to radiation generated lower down in the corona by collisions, which does enable some EUV detection out to as much as 6 \Rs \ \citep{Seaton2021, Auchere2023, DelZanna2025}. Still, the EUV emission at heights above 2 \Rs \ is exceptionally weak (requiring longer exposure times to observe) and is confined mostly to streamer tips. Further, the incoming radiation from lower down in the corona is highly angle-dependent and temporally variable \citep{Seaton2025}, making the emission difficult to use as a robust physical diagnostic. Thus, the strongly radiatively excited lines at longer wavelengths still offer an important contribution to the study of the solar corona, filling in parts of the corona beyond 1.5-2 \Rs \ that are difficult to probe with EUV alone.

\par
\begin{figure}[ht!]
\centering
\includegraphics[width= 0.9\columnwidth]{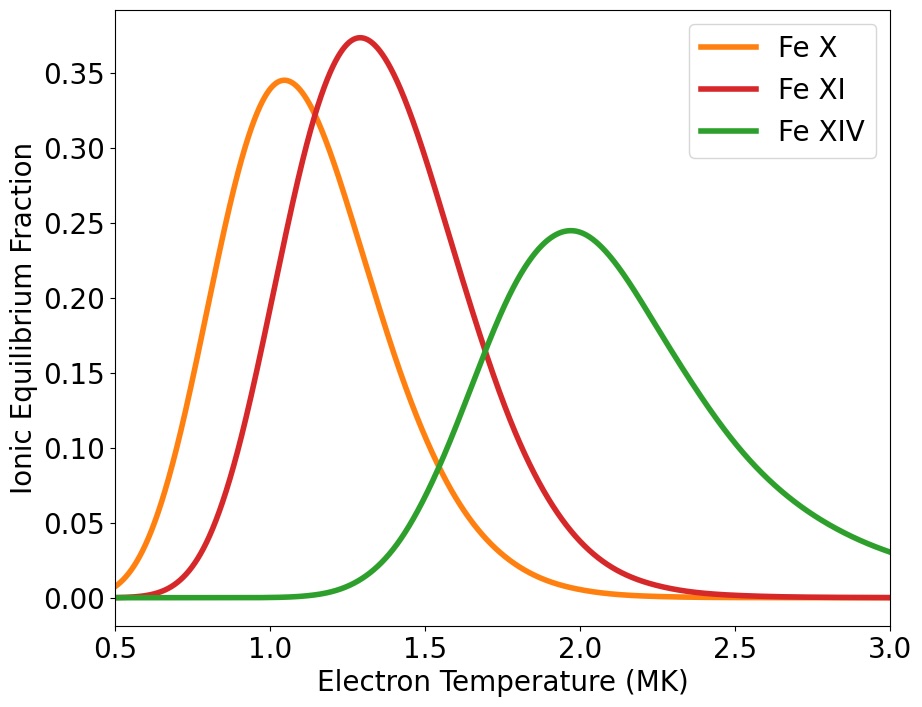}
\caption{The ionization equilibrium abundances of \ion[Fe x], \ion[Fe xi], and \ion[Fe xiv] as a function of $T_e$. Data are shown as spline interpolated values from the CHIANTI database \citep{Dere1997, Dufresne2024}.}
\label{fig:Ionization}
\end{figure}

Furthermore, a large variety of powerful diagnostics have been demonstrated in recent years to utilize these radiatively excited lines. Comparing the relative emission of these lines provides information on the electron temperature ($T_e$) distribution along the line-of-sight (LOS; e.g., \citealt{Habbal2010b, Habbal2021, Boe2023}), via the ionization equilibrium abundances of the different ions (as plotted in Figure \ref{fig:Ionization} for \ion[Fe x], \ion[Fe xi], and \ion[Fe xiv]). The widths of the lines can reveal non-thermal motions (e.g., \citealt{Mierla2008, Koutchmy2019, Muro2023, Zhu2024}) as well as the ion temperatures \citep{Esser1999}. The Doppler shift of the line center makes it possible to infer the bulk motions along the LOS (e.g., \citealt{Mierla2008, Zhu2024}), while the time variability in the Doppler shift and intensities can be used to infer the properties of coronal waves \citep{Tomczyk2007, Morton2025, Hahn2025}. Changes in the emission and Doppler signature of the lines also enable studies of the dynamics and propagation of Coronal Mass Ejections (CMEs; e.g., \citealt{Srivastava2000, Akmal2001, Ding2017, Boe2020a}), which is now being explored from space with the Visible Emission Line Coronagraph (VELC) on Aditya-L1 \citep{Ramesh2024, Priyal2025}. The polarization of the line profile can even provide constraints on the coronal magnetic field direction \citep{Tomczyk2008, Gibson2017} and strength (for lines in the infrared; \citealt{Lin2004, Schad2024b}). 

\par

Nonetheless, observing line emission at visible and infrared wavelengths also has its drawbacks (which are different than those for EUV lines), namely that coronal emission at those wavelengths is dominated by continuum scattering of the K-corona (Thompson scattering by electrons) and F-corona (interplanetary dust scattering, becomes the Zodiacal light at large elongation angles; see \citealt{Mann1998}). The K-corona itself is useful for measuring the coronal density, after the F-corona emission is removed. The K- and F-corona are typically separated from each other by a polarization method of white-light emission \citep{vanDeHulst1950}, but recent work demonstrated that they can also be separated using a color-based method (since the F-corona is red in color; \citealt{Boe2021}). 
\par

Comparing the line emission profiles as a function of helioprojective distance along structures to the corresponding continuum emission from the K-corona additionally facilitates a method for inferring the distance at which the ions freeze-in to a fixed ionization state as they flow outward into the solar wind \citep{Habbal2007, Habbal2011, Boe2018}. So even though the K- and F-corona brightness requires instrumental considerations to remove from the line emission, it does enable additional physical inferences which expand the potential of such data.
\par

Despite the unique advantages of emission lines at visible and near-infrared wavelengths, a majority of the observations at those wavelengths in recent years have focused on the continuum emission for the sake of inferring the structure and changes in the coronal density (e.g., the Wide-field Imager for Solar Probe, WISPR \citealt{Vourlidas2016}, and the Polarimeter to UNify the Corona and Heliosphere , PUNCH, \citealt{Deforest2022}). While that information is incredibly useful, particularly when there is time variability to study CMEs and other transient features, it does not provide information on the $T_e$ or elemental abundances, which are essential for answering questions about the heating of the corona and the exact formation processes of the solar wind. Even so, some information on $T_e$ can be extracted by narrowband polarized brightness observations of the K-corona around 400 nm (with the COronal Diagnostic EXperiment, CODEX, \citealt{Reginald2023}). Observations of the ionic emission over the same coronal region could, in combination with this method, be used to validate the calibration of such observations and could be used to distinguish any differences between the thermal and ionization temperature, which could be valuable for constraining non-equilibrium processes in the corona.

\par

Recent observations have also been pushing the capabilities of detecting visible and near-infrared emission out to larger heights in the corona. \cite{Boe2022} demonstrated emission from the visible and near-infrared lines of \ion[Fe xiv], \ion[Fe x], and \ion[Fe xi] out to as much as 3.4 \Rs \ during solar-minimum (at the end of Solar Cycle 24), making the extent comparable to the UVCS measurements of ionic line emission. In this work, we present these same three lines observed during the 2023 April 20 TSE, close to the maximum of Solar Cycle 25, continuously outward from just above the solar disk out to 6 \Rs \ -- probing ionic line emission throughout the entire middle-corona (from 1.5 to 6 \Rs, \citealt{West2023}) for the first time.
\par
In Section \ref{sec:Eclipse}, we discuss the eclipse expedition and describe the instrumentation deployed in Australia, while detailed descriptions of the data reduction, processing, and calibration are included in Appendix \ref{sec:Appendix}. In Section \ref{sec:Results}, we review the observations and initial findings of the line emission (Section \ref{sec:Lines} and \ref{sec:Ratio}), the corresponding continuum emission near each line (Section \ref{sec:Cont}). A discussion of the key findings can be found in Section \ref{sec:Conc}.

\section{The 2023 Total Solar Eclipse Expedition} 
\label{sec:Eclipse}

The 2023 April 20 TSE occurred during the rising phase of Solar Cycle 25 near the peak of solar activity. The eclipse was hybrid, with totality occurring in parts of Western Australia, East Timor, and parts of Indonesia. Our primary observing site was located in Exmouth, Australia, on the North West Cape (also referred to as the Ningaloo Peninsula), which had our entire complement of white-light and narrowband imaging systems. In Exmouth, totality lasted 53.8 seconds at 03:30 UT (11:30 local time). We had a secondary observing site, with more limited instrumentation comprised of only \ion[Fe xiv] and \ion[Fe xi] imaging systems, on a small island in the cluster of Lowendal Islands, approximately 208 km northeast of Exmouth. Totality on the island lasted for 61.9 seconds and occurred at 3:34 UTC (11:34 local time). The secondary site served as a safeguard against possible weather issues at Exmouth, but both sites had excellent observing conditions during totality (zero cloud cover), and so the data from both sites could be combined to increase the coronal signal-to-noise ratio (SNR). At the Exmouth (Island) site, the Sun was at a high altitude of 54.3 (56.3) degrees, creating ideal observing conditions despite the short duration of totality. In fact, the short totality meant that the Moon was nearly the exact same angular size as the Sun in the sky (eclipse magnitude of 1.00443), so almost all of the corona down to the chromosphere could be seen during the entire eclipse. 
\par

Indeed, the upper-chromosphere (and prominences) can be seen well in the simple slit-less flash spectrum of the corona taken with a 4K (3840x2160) color video camera (Panasonic HC-X1500, 600mm lens) through a transmission diffraction grating shown in the bottom right panel of Figure \ref{fig:Whitelight}. The slit-less spectrum shows the Balmer series of Hydrogen lines, the strong Helium line (the yellow 587.6 nm \ion[He I] line), as well as the green \ion[Fe xiv] line, representing a modern recreation of the original ``Coronium" detection by \cite{Young1872}. There is even some faint emission from what is likely the red \ion[Fe x] line (just to the left of the bright red H-$\alpha$ line).

\par
We discuss the eclipse instruments used to record the high spatial resolution white-light images in Section \ref{sec:Whitelight} and the narrowband line emission in Section \ref{sec:Narrow}. For all observations, we performed standard dark frame subtraction and flat fielding. Dark frames were taken immediately after totality by completely covering the telescopes with both lens caps and towels before running the same exposure sequences used during totality. Flat fields were taken by looking at clear blue sky near zenith (but away from the Sun to avoid direct illumination) with white translucent plastic coverings on top of the lens (i.e., the cover was out of focus and so recorded the relative optical efficiency throughout the system). The plastic coverings were then rotated in increments of 60 degrees and repeated six times (so every possible angle for the holders) in order to reduce any possible systematics introduced by the flat-fielding procedure. 

\subsection{White-light Imagers}
\label{sec:Whitelight}
High-resolution white-light imaging is useful for characterizing the fine-scale structures of the corona. For our observing site in Exmouth, we used a 1000mm lens with a Nikon Z6 II camera to observe the lower portion of the corona, out to about 3 \Rs. A composite image is shown in the top panel of Figure \ref{fig:Whitelight}, which is comprised of 77 unique exposures taken during totality. To probe the corona to larger helioprojective distances, we used two identical wide-field telescope systems composed of 200mm lenses with Nikon D810 cameras. For all the white-light telescopes, the exposure times spanned 1/125s up to 1s. A composite of these wider-field observations out to about 10 \Rs \ is shown in the lower left panel of Figure \ref{fig:Whitelight}, which is made from 199 images from the 200mm cameras as well as the 1000mm observations. The white-light data were aligned using a phase-correlation method as described in \cite{Druckmuller2009}, and processed to enhance fine-scale features by removing much of the low spatial-frequency signal in the overall drop of coronal brightness with distance from the Sun, as introduced by \cite{Druckmuller2006}.

\par
These white-light observations display the classic solar maximum shape of the corona (see \citealt{Maunder1899}), with coronal streamers occurring at almost all latitudes, intermixed with small open-field corridors (as opposed to a solar minimum corona with large equatorial streamers and polar coronal holes). A small CME was also present in the southwest region of the corona, seen in the wide field image at about 6-8 \Rs.

\subsection{Narrowband Imagers}
\label{sec:Narrow}

\begin{figure*}[ht!]
\centering
\includegraphics[width=0.975\linewidth]{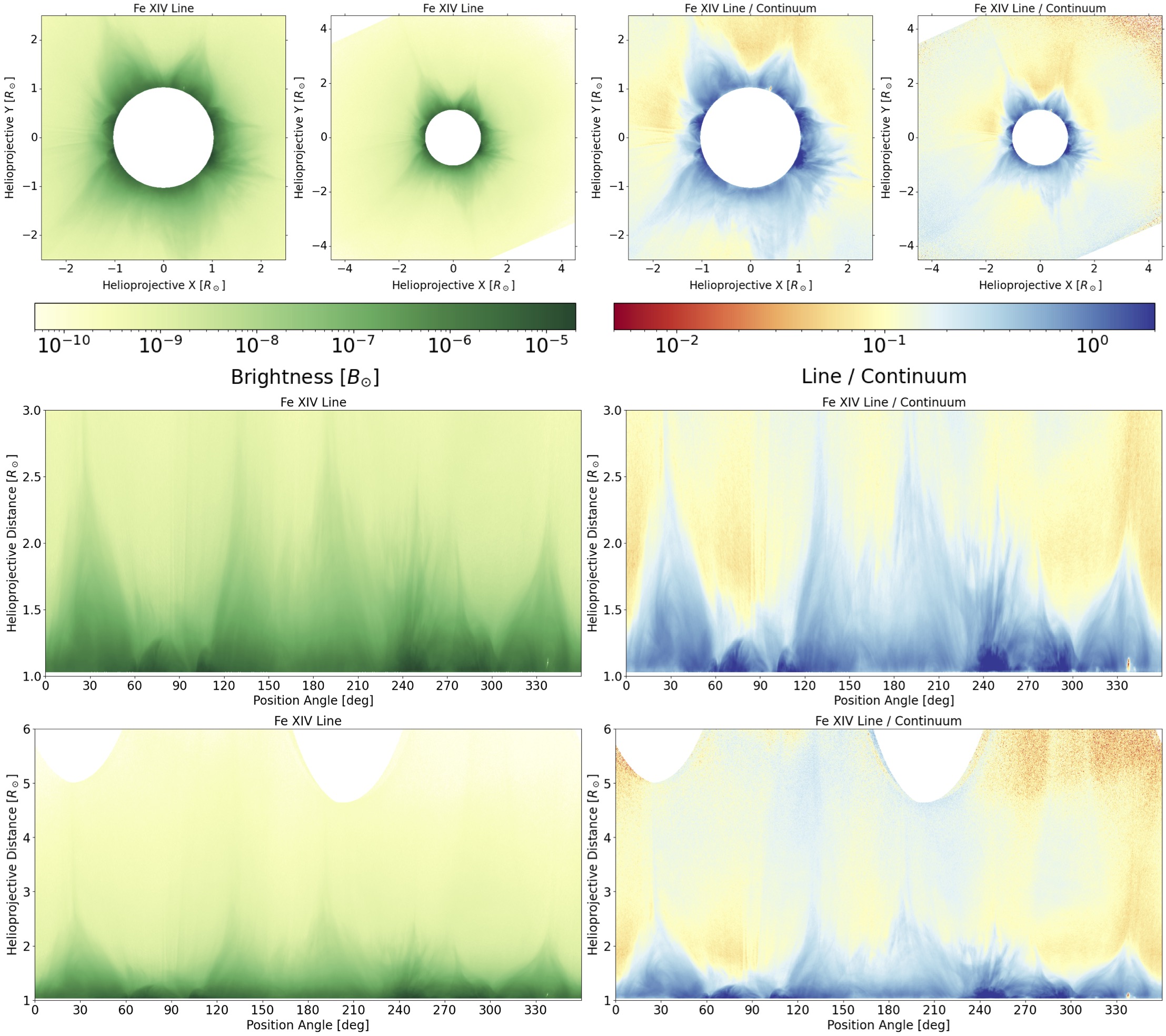}
\caption{Emission of \ion[Fe xiv] recorded with narrowband images (see Section \ref{sec:Narrow}). The panels on the left show the absolute line emission (in units of mean solar disk brightness) in log space, while the panels on the right show the line emission divided by the continuum (K+F corona) brightness near the same wavelength of the line. The top panels on each size show the corona with a window of 2.5 \Rs \ (left) and 4.5 \Rs \ (right) with Solar North upwards, while the bottom pairs of panels are in polar coordinates (Helioprojective distance vs Position Angle) extending out to 3 (top) and 6 \Rs (bottom).}
\label{fig:FeXIV}
\end{figure*}

\begin{figure*}[ht!]
\centering
\includegraphics[width=0.975\linewidth]{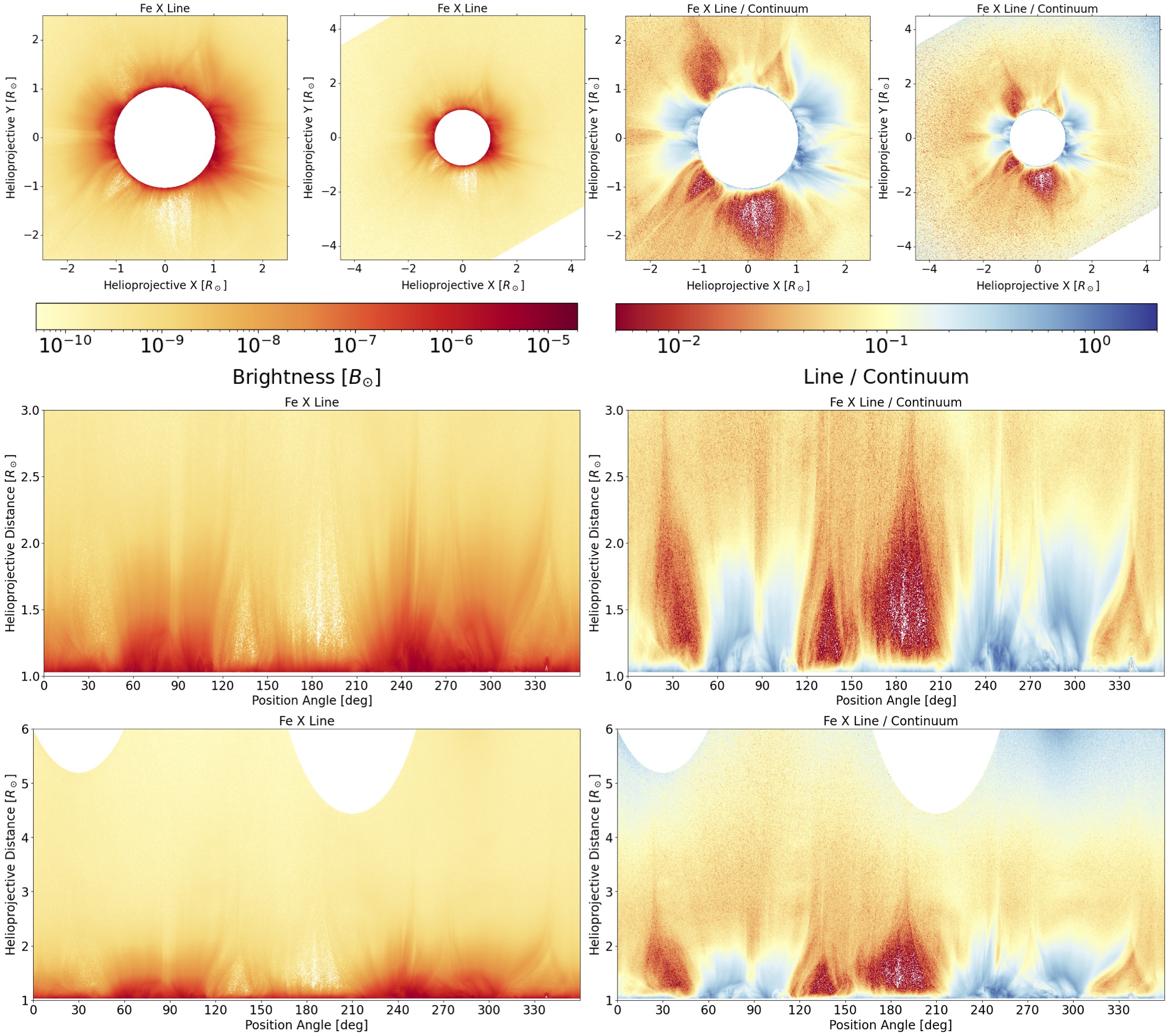}
\caption{Same format as Figure \ref{fig:FeXIV} for \ion[Fe x] emission.}
\label{fig:FeX}
\end{figure*}

\begin{figure*}[ht!]
\centering
\includegraphics[width=0.975\linewidth]{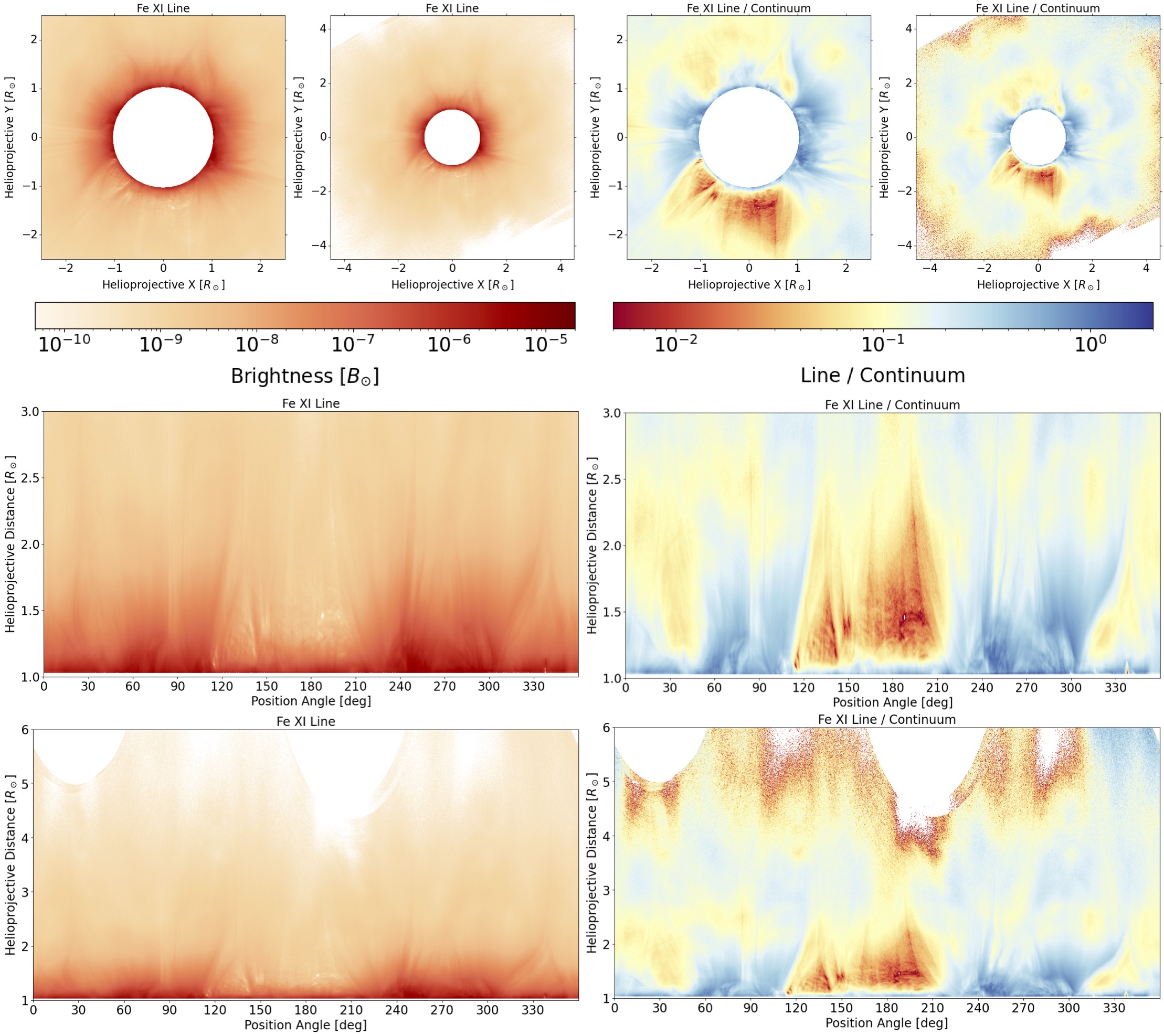}
\caption{Same format as Figures \ref{fig:FeXIV} and \ref{fig:FeX} for \ion[Fe xi] emission.}
\label{fig:FeXI}
\end{figure*}

To record the ionic emission lines, we performed observations with a suite of telescopes that were custom-made for this eclipse with modern high-resolution detectors and new narrowband filters. The observations of the line emission for \ion[Fe xiv] are shown in Figure \ref{fig:FeXIV}, \ion[Fe x] in Figure \ref{fig:FeX}, and \ion[Fe xi] in Figure \ref{fig:FeXI}. A qualitative composite of the \ion[Fe xiv] and \ion[Fe x] emission is shown in Figure \ref{fig:Whitelight}.

\par
The narrowband filters were manufactured by the Alluxa corporation with bandpasses with a full width at half maximum (FWHM) of approximately 1-1.5 nm (See Appendix \ref{sec:Filters} for the exact bandpasses). For each emission line, we used two telescopes -- one with a bandpass centered on top of the emission line, and a secondary one separated by $\approx 7$ nm towards shorter wavelengths (i.e., blueward) than the emission lines. The secondary ``Off-band" telescopes are essential since there is a strong continuum emission caused by scattering of photospheric light by electrons (K-corona; see \citealt{Lamy2020}) and dust throughout the solar system (F-corona; see \citealt{Morgan2007}). In fact, the K-corona and F-corona have a brightness similar to the emission lines, even as low as about 1.3 \Rs, and are the dominant source of emission at visible wavelengths beyond 2 \Rs \ at most (see Section \ref{sec:Cont}). 

\par

Each narrowband telescope consisted of a 300mm Zeiss f/4 telephoto lens (7.5 cm aperture) with a narrowband filter on the front and a ZWO astrophotography camera on the back. For \ion[Fe x] and \ion[Fe xiv] we used the ZWO ASI1600MM camera (Panasonic MN34230ALJ sensor), which has a resolution of 4656×3520 pixels, with a pixel size of 3.8 $\mu m$. For \ion[Fe xi] we used the ASI294MM camera (SONY IMX294 CMOS sensor), due to its higher sensitivity at infrared wavelengths. These cameras had a resolution of 8288x5644 pixels, with a pixel size of 2.3 $\mu m$, which was re-binned into 2x2 pixels for a resolution of 4144x2822. Unfortunately, an unforeseen consequence of using the ASI294MM camera was that the back-illuminated chip design resulted in an \'Etalon-type interference pattern that caused some minor contamination of the \ion[Fe xi] observations (only where the corona was very faint beyond about 3 \Rs). This issue did not appear during prior lab testing and is only visible with an extreme brightness contrast source such as that which occurs during a total solar eclipse.

\par

All of the narrowband camera systems cycled through exposure times during totality, ranging from 0.025 s up to 3.2 s, bracketed by factors of two for a total of 8 unique exposure times in the observing cycle. The sequence was then repeated throughout totality for a total of 32-35 usable exposures for each telescope system. Since totality was rather short, we opted not to do a simple exposure sequence in order, but rather had more of the shorter exposure times to coincide with the several seconds with the start and end of totality, and had additional longer exposures occur during the middle of totality. Optimizing the sequence in this way enabled more usable exposures without saturation contamination from the solar photosphere while also capturing as low down in the corona as possible during the shorter time that it was exposed.

\par
After the data reduction, all eclipse frames were co-aligned using a phase correlation method to account for the relative offset between each frame of the same telescope caused by the imperfect tracking by the mount. The data for each telescope were then stacked after accounting for the linearity of the cameras and photometrically calibrated using solar disk observations made after the eclipse (See Appendix \ref{sec:Calib} for more details). Once each stacked observation was compiled, they were all co-aligned with each of the other observations (at both Exmouth and the Island), including the offset, rotation, and relative scaling of the images. Finally, the bright prominence in the North west part of the low corona was used to align the images with Solar North via the Global Oscillation Network Group (GONG) H-$\alpha$ data \citep{Harvey1996, Hill2018}. The continuum, as measured by each off-band dataset, was then subtracted from the on-band emission to isolate the line emission observation. 
\par

\section{Results} 
\label{sec:Results}
\subsection{Line emission of \ion[Fe xiv], \ion[Fe x], and \ion[Fe xi]}
\label{sec:Lines}

\begin{figure*}[ht!]
\centering
\includegraphics[width=0.975\linewidth]{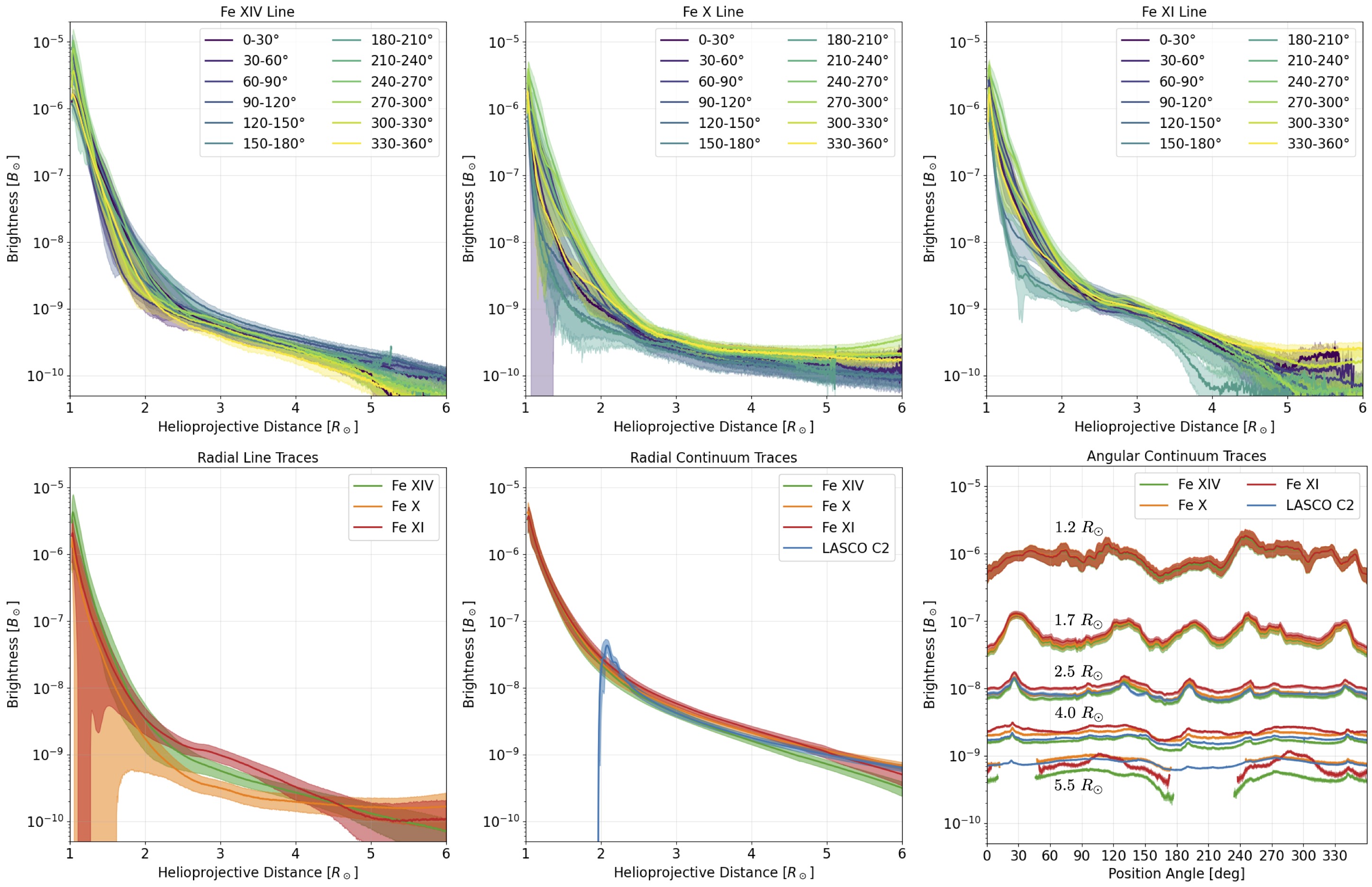}
\caption{Quantitative traces of the observed line and continuum emission. The top panels show radial traces of the line brightnesses averaged over 30-degree position angle windows for \ion[Fe xiv] (left), \ion[Fe x] (middle), and \ion[Fe xi] (right). The lines represent the median average, while the bands represent the standard deviation of the data in each region (which was always larger than the photometric error). The bottom left panel shows the brightness of all of the lines as a function of helioprojective distance, averaged over all position angles. The bottom middle panel is the same as the bottom left, but for the continuum observations corresponding to each line, as well as LASCO-C2. The bottom right panel shows position angle traces of those same continuum observations at a distance of 1.2, 1.7, 2.5, 4.0, and 5.5 \Rs, averaged in a window of $\pm 0.05$ \Rs for each.}
\label{fig:Traces}
\end{figure*}

Upon completion of the reduction and processing described in Section \ref{sec:Eclipse} (and Appendix \ref{sec:Appendix}), we produced observations of the emission from the \ion[Fe xiv], \ion[Fe x], and \ion[Fe xi] lines from just above the solar surface (about 1.03 \Rs) outward to 6 \Rs, as shown in Figures \ref{fig:FeXIV}, \ref{fig:FeX}, and \ref{fig:FeXI} respectively. Quantitative radial traces of the line emission are shown in Figure \ref{fig:Traces}. 

\par
Due to the relatively different ionization equilibrium abundances as a function of the $T_e$, which are plotted in Figure \ref{fig:Ionization}, each of these emission lines originates from distinct structures along each LOS. Indeed, since the corona is optically thin, we expect to see a range of structures along the LOS with different $T_e$, which can be resolved by comparing the emission of these different ionic lines (e.g., \citealt{Landi1997, Habbal2010a, Boe2023}). The emission from each line can be thought of as a slice of the corona over the range of temperatures that the ion can exist, weighted by the density of the corona in those structures. Consequently, all of the lines fade dramatically with distance from the Sun as the density in the corona drops. Displaying the lines relative to the continuum (K+F corona) then offers a simple method to visualize the relative emission of the lines scaled by the background electron density. While this observable will become biased by F-corona (dust) brightness farther out, it still enhances the visualization of the line.

\par
For the \ion[Fe xiv] line, which is weighted towards plasma from about 1.6 to 2.4 MK, the emission is brightest in closed field lines where the $T_e$ is expected to be higher. In the data shown in Figure \ref{fig:FeXIV}, the \ion[Fe xiv] emission is significant out to a few solar radii in the streamer structures, and is very strong in smaller closed field lines near the equator (up to $2 \times 10^{-5} \ B_\odot$). Once the coronal field becomes more open, the \ion[Fe xiv] emission drops significantly. In the low closed structures, the line emission is comparable to or even a few times brighter than the background continuum, but it quickly drops to about 10$\%$ of the continuum brightness farther out in the corona. However, the line emission then plateaus and remains around 15-20$\%$ of the continuum brightness outward to 6 \Rs. 

\par
In contrast to the \ion[Fe xiv] line, the \ion[Fe x] line is weighted towards cooler plasma from about 0.7 to 1.4 MK. Hence, the \ion[Fe x] line originates primarily from open-field regions such as coronal holes. In Figure \ref{fig:FeX}, the \ion[Fe x] emission is brightest in the open-field corridors seen near the equator of the Sun (around 50-110 and 220-310 PA) with values of a few times $10^{-6} \ B_\odot$. On the other hand, the \ion[Fe x] emission is extremely faint in the hotter closed field regions near the solar poles. In fact, the southern streamers (from about 115-215 PA) have a nearly undetectable amount of \ion[Fe x] emission, particularly at a distance of 1.2 to 2 \Rs. The \ion[Fe x] emission is likely still present in those streamers, but the SNR is simply too low to extract it from the background continuum signal in our dataset. Once the continuum brightness drops, the \ion[Fe x] emission returns and is seen across all position angles out to 6 \Rs, with values around a few times $10^{-10} \ B_\odot$. The \ion[Fe x] brightness has a similar relationship to the continuum as does \ion[Fe xiv], with the brighter regions getting close to the same brightness as the continuum, but then dropping to about 10 $\%$ of the continuum brightness.

\par
The \ion[Fe xi] emission, which is weighted towards plasma from about 1.0 to 1.8 MK, unsurprisingly shows a similar behavior to \ion[Fe x]. The slightly higher temperature response of \ion[Fe xi] does lead to the line being stronger in the southern streamers, though it too has a few points where the emission is not detectable above the SNR. However, the \ion[Fe xi] shows a lot of fine-scale structures throughout the streamers as well as in the open-field corridors. In contrast with the other two lines, it provides an intermediate scan of coronal temperatures. 
\par
Unfortunately, the detectors for \ion[Fe xi] had an \'Etalon-type interference issue originating from the back-illuminated detector (and near-infrared wavelength) which contributed some contamination of the emission beyond around 2.5-3 \Rs. While the contamination blurs the exact spatial behavior of the emission, it is still apparent that the line is strong out to at least 4 \Rs \ with likely detections out to 6 \Rs (albeit with a lower confidence than \ion[Fe x] and \ion[Fe xiv]). In fact, around 2.5 to 4.5 \Rs, the \ion[Fe xi] line is the brightest of the three lines.

\par
Despite the spatial variance across unique fine-scale structures, all three lines show a similar behavior with distance. These lines are strong low down in the corona (at least $4 \times 10^{-6} B_\odot$, with \ion[Fe xiv] up to $2 \times 10^{-5} B_\odot$) and fade to about 1 -- 5 $\times 10^{-10} B_\odot$ at 6 \Rs \ for all position angles. All of the lines also show a rapid decrease in brightness between 1 and about 2 -- 2.2 \Rs, then the slope of the brightness drop bends to be much less steep beyond that distance (see Figure \ref{fig:Traces}). These kinks in the emission profiles are likely caused by the transition from the lower part of the corona, where collisional excitation of the lines is common, to the outer corona, where collisions fade and radiative excitation becomes the dominant line formation mechanism (see \citealt{Habbal2007, Seaton2021, Boe2023}).

\subsection{Line Emission Ratios}
\label{sec:Ratio}

As discussed earlier in Section \ref{sec:Lines}, the exact brightness of each emission line depends on the density and $T_e$ of each structure. Thus, the relative emission of the lines can be used as a probe of the physical state of the corona. All lines show a similar trend in the structure of the line emission with heliprojective distance, which is caused by the steep drop-off in coronal density. To highlight the differences in the $T_e$ of various coronal structures, we took the ratio of each line to every other line (which largely corrects for the overall density). 

\par

\begin{figure*}[ht!]
\centering
\includegraphics[width=0.975\linewidth]{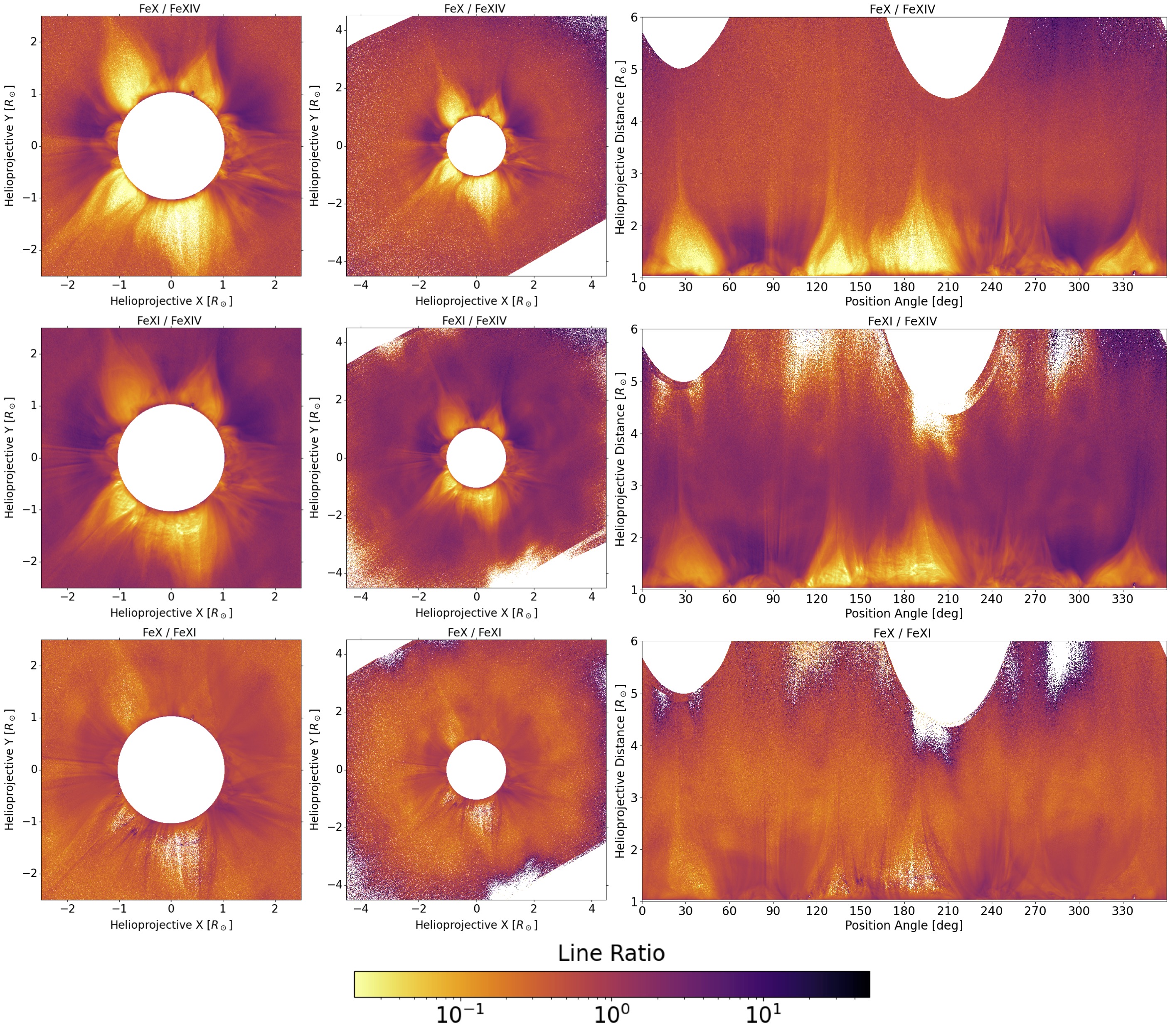}
\caption{Collection of line emission ratios of \ion[Fe x] / \ion[Fe xiv] (top row), \ion[Fe xi] / \ion[Fe xiv] (middle row), and \ion[Fe x] / \ion[Fe xi] (bottom row). The emission ratios are shown out to 2.5 \Rs \ in the left panels and 4.5 \Rs \ in the middle panels, and in polar coordinates out to 6 \Rs in the right panels.} 
\label{fig:LineRatios}
\end{figure*}

The line emission ratios of \ion[Fe x]/\ion[Fe xiv], \ion[Fe xi]/\ion[Fe xiv], and \ion[Fe x]/\ion[Fe xi], are all shown in Figure \ref{fig:LineRatios} in the top, middle and bottom rows respectively. The left column shows these ratios out to 2.5 \Rs, showcasing the variability of the $T_e$ of structures in the lower corona. The middle column shows the ratios out to 4.5 \Rs, and the right column shows the ratios in polar coordinates, which probes the $T_e$ structures out to 6 \Rs.

\par

It is important to note that these line ratios really represent the emissivity-weighted LOS $T_e$ distribution. While they do not provide an exact $T_e$ for a specific chunk of coronal plasma, they do represent a physically interesting value illustrating the temperature structure of the corona. The combination of all three lines could then be used to infer the $T_e$ distribution along each LOS, after accounting for the excitation mechanisms of each line. This sort of inference is referred to as a Differential Emission Measure (DEM, see \citealt{DelZanna2018}), though DEMs have typically been done with collisionally excited EUV and X-ray lines. A Radiative version of a DEM (RDEM) was introduced by \cite{Boe2023} and applied to these same three emission lines using data from the 2019 TSE, leveraging a 3-dimensional (3D) global Magnetohydrodynamic (MHD) simulation to account for the relative amount of collisional and radiative excitation. We also intend to produce such inferences with these data, including a 3D MHD model to estimate collisional excitation of the lines. Yet, the MHD simulation would itself warrant significant analysis, discussion, and comparison to other observational datasets. Therefore, a full RDEM inference is beyond the scope of this initial instrumentation, calibration, and observation paper. Here, we will discuss the initial findings from these line ratios.

\par

Based on the emission line ratios, the corona is divided into two main structures: a few large higher-temperature polar streamers and two large lower-temperature equatorial open-field regions. The highest temperature regions of the corona during this eclipse are clearly the two southern streamer arcades between about 120 and 215 PA. Inside these streamers, \ion[Fe xiv] emission dominates, while the \ion[Fe x] emission drops below the detection threshold in the streamer cores. The \ion[Fe xiv]/\ion[Fe xi] line ratio highlights how those streamers have by far the largest relative amount of \ion[Fe xiv] emission, indicating a $T_e$ of approximately 1.7-2 MK or more. 
\par

The streamers to the northeast (about 0 to 60 PA) and the northwest (300 to 360 PA) have considerably more \ion[Fe xi] and \ion[Fe x] emission relative to \ion[Fe xiv], indicating a more moderate average temperature closer to 1.5 MK. It is interesting that the northwest streamer, which has a large prominence, actually is much brighter in \ion[Fe x] emission compared to the other large streamers, suggesting a lower average $T_e$. However, there is a large open-field plume that is pushing on the streamer from the south (around 320 PA), which could be overlapping along the LOS and biasing the ratio towards the cooler ions.

\par
The other primary structures are two large open-field regions on the solar equator (at around 35-125 and 270-320 PA). These open-field corridors show a much larger amount of \ion[Fe x] and \ion[Fe xi] emission, suggesting a lower temperature around 1.1-1.4 MK. The \ion[Fe x]/\ion[Fe xi] ratio is weighted towards these cooler open-field lines, and so can resolve fine-scale plumes inside the open-field regions while not highlighting the higher $T_e$ streamers. The open-field corridors show a strong expansion with helioprojective height, as opposed to the streamers, which pinch off into narrow streamer stalks. The tips of these streamer stalks can be seen in some cases out to 6 \Rs in the line \ion[Fe xiv]/\ion[Fe x] ratio.

\par
These line ratios also change with helioprojective height throughout the corona, particularly showing more \ion[Fe x] and \ion[Fe xi] emission (below 1.3 \Rs) compared to \ion[Fe xiv] even at the base of hot streamers, but then an increasing amount of \ion[Fe xiv] beyond at larger heights. This finding was seen with these same lines for the 2019 eclipse \citep{Boe2023} and might be an indication that heating is occurring low down in the corona, or perhaps could be an effect related to the temperature dependence of the scale height (see \citealt{Aschwanden2000}). 

\par
There is also an intricate structure around 215 to 290 position angle that is not easy to define cleanly as either an open-field region or a streamer. It could be a mix of both based on the fact that it shows the brightest emission from all three lines. There was a similar report by \cite{Boe2020a}, who found that above an active region on the limb during the 2017 TSE had strong emission from \ion[Fe xi] and \ion[Fe xiv]. 

\par

Below about 2.5 to 3 \Rs, these line ratios (as well as the line emission itself) show a large amount of structural variation along the POS. That is, there is a preponderance of small-scale structures with unique temperatures. However, beyond 3 \Rs, the corona becomes increasingly flat regardless of the underlying structure. The \ion[Fe xi] emission (and ratios including it), unfortunately suffered from the \'Etalon-type interference issue (see Section \ref{sec:Narrow}), which limits their substantive analysis beyond 4 \Rs. Still, the \ion[Fe xiv]/\ion[Fe x] ratio becomes exceptionally flat throughout the corona, besides a few streamer stalks. This finding indicates that most of the corona is becoming nearly isothermal beyond 3 \Rs, which is consistent with work from prior eclipses as well as in situ data on the ionic abundances in the solar wind, independent of the solar cycle \citep{Habbal2010c, Habbal2021}. 

\par

Nevertheless, there is an interesting feature around a PA of 270 to 330 degrees, where the \ion[Fe x] emission rises and the \ion[Fe xiv] emission fades. The resulting line ratio indicates a significant deviation from the rest of the corona. This region corresponds to a location directly above a large open-field corridor that has strong \ion[Fe x] emission lower down in the corona. It is plausible that this increase is due to time-dependent release of cooler material from that open-field region (or a tiny CME), perhaps through some sort of interchange reconnection with the streamers on either side. However, it is difficult to make any strong claims about this finding given the low SNR of the emission in that region.

\subsection{Continuum Observations}
\label{sec:Cont}

\begin{figure*}[ht!]
\centering
\includegraphics[width=0.975\linewidth]{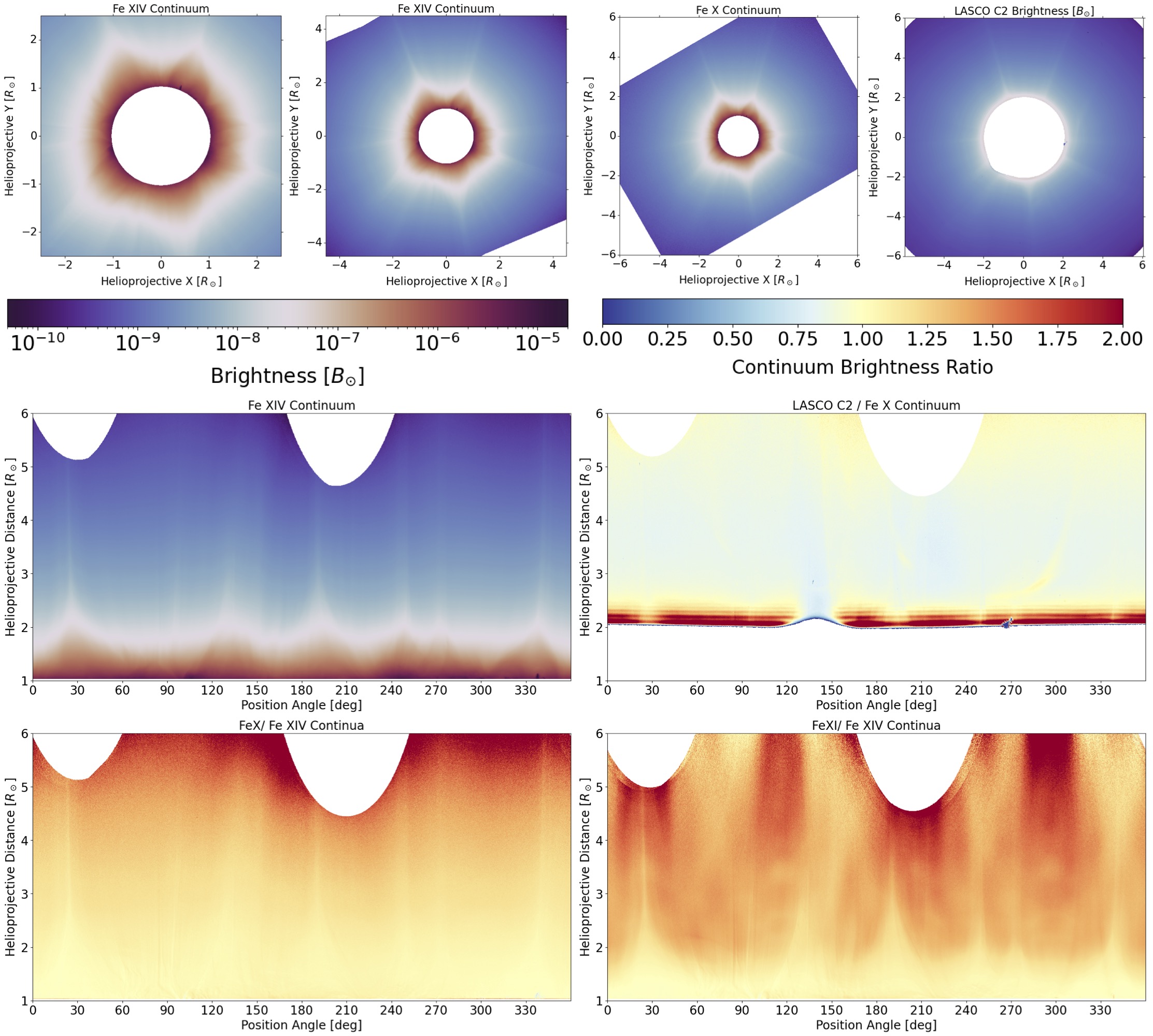}
\caption{Observations of the continuum emission during the eclipse (see Section \ref{sec:Cont}). The top row shows the continuum near the \ion[Fe xiv] line out to 2.5 (left) and 4.5 (middle-left), the \ion[Fe x] continuum out to 6 \Rs (middle-right) as well as the LASCO-C2 emission out to 6 \Rs (right). The bottom panels then show the \ion[Fe xiv] continuum in polar coordinates out to 6 \Rs (top left), as well as continua ratios of LASCO-C2/\ion[Fe x] (top right), \ion[Fe x]/\ion[Fe xiv] (bottom left), and \ion[Fe xi]/\ion[Fe xiv] (bottom right)}.
\label{fig:Cont}
\end{figure*}

The continuum observations from all off-band narrowband observations (see Section \ref{sec:Eclipse}), as well as the LASCO-C2 observations at the exact same time as totality in Exmouth (03:30 UT), are shown in Figure \ref{fig:Cont}. Quantitative traces of the continuum data are shown in Figure \ref{fig:Traces}, where the bottom middle panel contains traces as a function of helioprojective distance and the bottom right panel has traces at fixed heights of 1.2, 1.7, 2.5, 4.0, and 5.5 \Rs \ (each averaged over $\pm 0.05 $ \Rs) for all position angles. 
 
\par

The continuum observations show the combination of both the K-corona (electron scattering) and F-corona (dust scattering). As discussed in Sections \ref{sec:Intro} and \ref{sec:Eclipse}, the F-corona has a color relative to the solar spectrum, where it is brighter at redder wavelengths. This color effect necessitates the deployment of off-band telescopes with narrowband filters near each line. These observations can then provide information on the K and F corona via their relative brightness. 
\par

The continuum data for the \ion[Fe x] off-band (at 630.4 nm, see Appendix \ref{sec:Filters}) are used as the primary reference for the absolute calibration of all the imaging data (using solar disk observations, see Appendix \ref{sec:Calib}). However, the \ion[Fe xiv] continuum has the lowest F-corona contribution, so it is the best approximation of the K-corona. The \ion[Fe xiv] continuum data are shown out to 2.5 and 4.5 in the top left two panels of Figure \ref{fig:Cont}. The \ion[Fe x] continuum is the closest in wavelength to the LASCO-C2 Orange filter (see \citealt{Lamy2020}), so we compare the \ion[Fe x] continuum and LASCO-C2 observations with the same perspective of 6 \Rs \ in the top right panels. The \ion[Fe xiv] continuum is then shown in polar coordinates in the middle-left panel. The ratio of the brightness of LASCO-C2 and the \ion[Fe x] continuum, as well as the \ion[Fe x] and \ion[Fe xi] continua relative to the \ion[Fe xiv] continuum, are shown in the middle-right, bottom-left, and bottom-right panels, respectively. 
\par

The overall structure of the continuum is rather simple, with the streamers at the north and south having the largest brightness and extending outward with stalks going out to at least 6 \Rs. The regions identified as cooler open-field corridors in Section \ref{sec:Ratio} do have a lower continuum brightness, but they are not nearly as easy to distinguish as they were with the line emission data. In general, this corona near solar maximum has a remarkably small variation in continuum brightness at any given helioprojective distance, other than the small streamer stalks and small closed field regions low down (below about 1.5 \Rs).
\par

Farther out in the corona (beyond 2 \Rs), the observed LASCO-C2 emission is quite consistent with our \ion[Fe x] continuum. These LASCO-C2 data were taken from the Legacy Archive\footnote{ \url{http://idoc-lasco.ias.u-psud.fr/sitools/client-portal/doc/}}, which provides processed data accounting for stray light, vignetting, cosmic rays, orientation of the spacecraft, and a photometric calibration from standard stars \citep{Llebaria2012SPIE, Lamy2014, Pagot2014, Lamy2020}. 
\par

The independent comparison to LASCO-C2 validates our photometric calibration, matching exactly at a height of about 2.5 \Rs. The LASCO-C2 brightness is considerably higher below 2.5 \Rs (above the occulter), likely indicating the presence of additional light from the solar disk that has diffracted around the occulter. There are also a few streaks of stray light present around 70, 190, and 250-320 PA at distances of roughly 2.8-4.5 \Rs. 

\par
The brightness of the LASCO-C2 brightness falls after 2.5 \Rs, relative to our continuum observation. It then becomes closer to our data again, beyond about 5 \Rs. This behavior hints at a slightly inaccurate accounting of the LASCO-C2 vignetting profile as a function of height. Further, the occulter arm, at about 130-150 PA, shows the lowest relative continuum brightness, strongly indicating that the attenuation caused by the blocking arm is not perfectly accounted for. Nevertheless, the LASCO-C2 and eclipse continuum agree to within 10-20$\%$ throughout the entire corona, other than near the occulter and the occulter arm. This comparison highlights the exceptional accuracy that has been achieved with the LASCO-C2 processing and showcases how valuable eclipse data are for validating coronal observations by coronagraphs and vice versa.

\par

The continuum observations for the \ion[Fe x] and \ion[Fe xi] lines are shown as a ratio to the \ion[Fe xiv] continuum in the bottom panels of Figure \ref{fig:Cont}. As previously demonstrated by \cite{Boe2021}, the F-corona has a color effect where it becomes brighter at longer wavelengths. This effect causes the \ion[Fe x] continuum to slowly rise relative to \ion[Fe xiv] continuum, becoming more than 50$\%$ at about 4.5-5 \Rs. The \ion[Fe xi] continuum shows an even more dramatic effect, becoming 50$\%$ brighter than the \ion[Fe xiv] continuum as low as 1.5 \Rs. Still, all continua observations agree below about 1.3 \Rs.

\par
The combination of these continuum data can be used to extract the relative K- and F-corona signals, but to do so robustly requires an accounting of the slight color that the K-corona has due to limb-darkening. Limb-darkening changes at different wavelengths, which in turn changes the angle-dependent incident radiation that changes the K-corona brightness (also depending on the 3D density structure of the corona). The best way to account for this effect is with a global 3D MHD model, which we intend to do in the future. Ergo, we defer that more detailed analysis to a subsequent study.

\section{Discussion and Conclusions} 
\label{sec:Conc}
In this work, we presented observations from the 2023 TSE in Australia, including high-resolution white-light data (see Section \ref{sec:Whitelight}) and narrowband emission observations of the ionic emission lines of \ion[Fe xiv], \ion[Fe x], and \ion[Fe xi] (see Section \ref{sec:Narrow}). The narrowband data were carefully processed and calibrated to the solar disk (see Appendix \ref{sec:Appendix}). The narrowband continuum observations (see Section \ref{sec:Cont}), which were used to remove the background K- and F-corona continuum emission, were then validated against LASCO-C2 observations at the same time, demonstrating the reliability of our telescopic systems. The primary result of this work is the demonstration that visible and near-infrared line emission can be observed throughout the solar corona from just above the photosphere (1.03 \Rs) out to at least 6 \Rs \ (see Section \ref{sec:Lines}). 
\par
Additionally, these observations are possible with less than a single minute of exposure time, in contrast to prior observations of ionic line emission in the UV beyond 4 \Rs, which have typically required long exposure times of 10 minutes or even hours of co-added exposures to get reasonable SNR at these distances (e.g., \citealt{Zangrilli2012, Giordano2013, Auchere2023}). The main benefit to observing emission lines at visible and infrared wavelengths (and near-UV) is their ubiquitous excitation by photospheric light, hence they are an important tool for probing the corona beyond 1.5-2.5 \Rs.
\par

The nature of the line emission itself showcased some notable features of the corona during the eclipse, near the maximum of the solar activity cycle. Low down in the corona, there are a multitude of fine-scale structures seen both in the white-light image as well as in the structure of the line emission. There are plumes of open field lines emitting more \ion[Fe x] in contrast to multiple streamers that emitted much more \ion[Fe xiv] due to their higher $T_e$. The average $T_e$ can be qualitatively inferred from the line ratios (see Section \ref{sec:Ratio} and Figure \ref{fig:LineRatios}), which indicate that throughout the streamers and open field corridors, there was a large amount of spatial variation of the ionic lines, indicating variability in the density and temperature of the coronal structures. 
\par

On the other hand, beyond about 3 \Rs, the corona appears to become remarkably isothermal, with a nearly flat \ion[Fe x]/\ion[Fe xiv] line emission ratio throughout the corona. This behavior of a nearly isothermal distribution of temperature beyond about 2-3 \Rs \ has been seen commonly before during eclipse observations \citep{Habbal2021}, including during solar minimum \citep{Boe2023}, and based on the frozen ionic charge states seen in situ \citep{Habbal2010c}. This study further supports the consistency of these findings, even during solar maximum. It is important to note that the ions may be frozen-in by 3 \Rs (e.g., see \citealt{Boe2018}), meaning that the ``isothermal" temperature may apply not at 3 \Rs in particular, but rather to the freezing-in point of the majority of solar wind sources -- which is still a physically insightful finding for models to benchmark against. 

\par
The large array of valuable physical inferences that are possible with visible and infrared coronal line emission (see Section \ref{sec:Intro}), combined with the proven spatial extent of these radiatively excited emission lines at visible and infrared wavelengths out to at least 6 \Rs, as demonstrated in this work, emphasizes the importance of observing these lines farther out in the corona. In particular, it supports the necessity of future eclipse observations and the development of ground- and space-based coronagraphs to observe these lines. There are a number of ground- and space-based observatories recently coming into operation (or soon to be), yet none of them are capable of probing more than one of these lines throughout the middle corona.

\par
On the ground, all observations are currently limited in helioprojective extent below about 1.5 to 2 \Rs \ due to stray light and atmospheric brightness on the Earth. In particular, the Coronal Multi-channel Polarimeter (CoMP; \citealt{Tomczyk2008}) established routine observations of \ion[Fe xiii] 1074.7 nm and is now being replaced by an upgraded COMP (UCoMP; \citealt{Tomczyk2022} \footnote{\url{https://www2.hao.ucar.edu/mlso/instruments/upgraded-coronal-multi-channel-polarimeter}}). There have also been new high-spatial/spectro-polarimetric resolution observations by the Daniel K. Inouye Solar Telescope (DKIST; \citealt{Rimmele2020}). However, at present, DKIST can only observe the density-sensitive \ion[Fe xiii] line pair (1074.7 nm and 1079.8 nm) along with \ion[Si x] 1430 nm line over a very small field-of-view \citep{Schad2024a}, which does limit its current capabilities for probing the $T_e$ distribution in the corona.

\par

In space, there are new observations from Metis and from ASPIICS on PROBA-3 \citep{Galano2018, Zhukov2025} which will be able to probe the region between $\approx$1.1 and 4 \Rs \ (depending on orbital distance for each spacecraft). However, they will collectively only have filters for \ion[He ii] 30.4 nm, \ion[H i] 121.6 nm, \ion[He i] 587.7 nm, and \ion[Fe xiv] 530.3 nm. The one \ion[Fe xiv] line from ASPIICS will be a useful dataset for demonstrating time variability of the line, and for comparison to eclipse data, but it will be unable to measure the coronal $T_e$ on its own without another ionic line. Indeed, other than \ion[Fe xiv] with ASPIICS and VELC, no visible or infrared coronal emission line has been observed in space since the untimely failure of the LASCO-C1 instrument in 1998 (see Section \ref{sec:Intro}).

\par
Given the clear importance of these visible and infrared emission lines for better characterizing the physics of the solar corona beyond 2 \Rs, which is crucial for investigating the connection between the corona and solar wind, there is a strong need for a new generation of spacecraft that can observe several of these lines to large helioprojective distances (see the Decadal Survey White Paper by \citealt{BoeWP}). Occultation of the photosphere could be accomplished with a traditional coronagraph, a secondary spacecraft (recently proven by ASPIICS; \citealt{Zhukov2025}), or even using the Moon from a lunar orbit \citep{Habbal2013, LunarSOX2023}. Additionally, recent airborne TSE observations by \cite{Samra2022a, Samra2022b, Samra2025} have demonstrated that there are several emission lines farther in the infrared that could be used as well, further justifying the need for spectroscopic observations (via narrowband filters and/or spectrometers) spanning the visible and infrared spectrum.

\par

Just as observations of ionic emission at visible wavelengths launched the field of coronal physics over a hundred and fifty years ago, so too are these lines important for pushing the field forward now. They offer an important alternative that has distinctly different advantages and drawbacks compared to lines in the EUV and other wavelengths. A new generation of observatories, with complementary data spanning the entire wavelength range of corona emission, including X-rays, EUV, near-UV, visible, infrared, and radio (which provides important constraints on the magnetic field; see \citealt{Gary2013, Gary2023}), should be developed and deployed. Each of these wavelength bands offers unique information on the physics of the corona, and equipped with high-quality data from all of them simultaneously, we would likely be able to significantly advance coronal physics beyond what is possible with any one set of wavelengths alone.

\begin{acknowledgments}
We thank Adalbert Ding and Judd Johnson for their work developing earlier generations of narrowband instrumentation that led to these new instruments, as well as for their input and advice on the 2023 eclipse expedition.
\par
The eclipse expedition was supported by NSF grant AGS 21303171 (PI was SH). 
BB was supported by NSF SHINE grant AGS 2501212. 
This publication was supported by the project ``Innovative Technologies for Smart Low Emission Mobilities", funded as project No. CZ.02.01.01/00/23\_020/0008528 by Programme Johannes Amos Comenius, call Intersectoral cooperation.
SC was supported by a DKIST Ambassador fellowship (AURA N00032456C)

\par
This work makes use of the LASCO-C2 legacy archive data produced by the LASCO-C2 team at the Laboratoire d’Astrophysique de Marseille and the Laboratoire Atmosphères, Milieux, Observations Spatiales, both funded by the Centre National d’Etudes Spatiales (CNES). LASCO was built by a consortium of the Naval Research Laboratory, USA, the Laboratoire d’Astrophysique de Marseille (formerly Laboratoire d’Astronomie Spatiale), France, the Max-Planck-Institut für Sonnensystemforschung (formerly Max Planck Institute für Aeronomie), Germany, and the School of Physics and Astronomy, University of Birmingham, UK. SOHO is a project of international cooperation between ESA and NASA.
\par
This work utilizes GONG data obtained by the NSO Integrated Synoptic Program, managed by the National Solar Observatory, which is operated by the Association of Universities for Research in Astronomy (AURA), Inc. under a cooperative agreement with the National Science Foundation and with contribution from the National Oceanic and Atmospheric Administration. The GONG network of instruments is hosted by the Big Bear Solar Observatory, High Altitude Observatory, Learmonth Solar Observatory, Udaipur Solar Observatory, Instituto de Astrofísica de Canarias, and Cerro Tololo Interamerican Observatory.

\end{acknowledgments}

\begin{contribution}

BB was the primary author of the paper and led the work to reduce, calibrate, and analyze the narrowband data. BB also directed the narrowband observations on-site in Exmouth, including operating the narrowband imaging mount during totality. 
SH led the overall eclipse expedition and contributed significantly to the writing of the paper. 
MD processed the broadband white-light data and made the narrowband composite image in Figure \ref{fig:Whitelight}. MD also contributed significantly to the planning of the expedition. 
PS and MS were the primary designers and manufacturers of both the physical and software infrastructure to operate the instrumentation. 
PS operated the white-light and slit-less flash spectrum instrumentation during totality. 
MS and JH were the sole operators at the secondary Island site. 
EA, SC, and DC all contributed to the preparations in Exmouth, including transport and assembly of the equipment, and collection of both eclipse and calibration data. 
Contributions to the 3D printing of components was made by PS, MS, EA, and SC. 
SC assisted BB after the expedition in collecting lab calibration data.

\end{contribution}

\facilities{GONG, SOHO/LASCO-C2}

\appendix
%\vspace{-6mm}
\setcounter{figure}{0}
\renewcommand{\thefigure}{A\arabic{figure}}
\section{Instrument Characterization and Calibration}
\label{sec:Appendix}
This appendix contains an overview of details pertaining to the operation, collection, and analysis of the narrowband data from the 2023 April 20 TSE. The primary information about the expedition and types of data collected was described in Section \ref{sec:Eclipse}. On site in Exmouth, we had a complement of several individual telescopes to collect emission from the \ion[Fe xiv], \ion[Fe x], and \ion[Fe xi] emission lines and nearby coronal continuum. We then had a secondary site on an Island off the coast of Australia that had only \ion[Fe xiv] and \ion[Fe xi] imaging systems.

\par

\subsection{Instrumentation}
\label{sec:Inst}

\begin{figure*}[t!]
\centering
\includegraphics[width=0.975\linewidth]{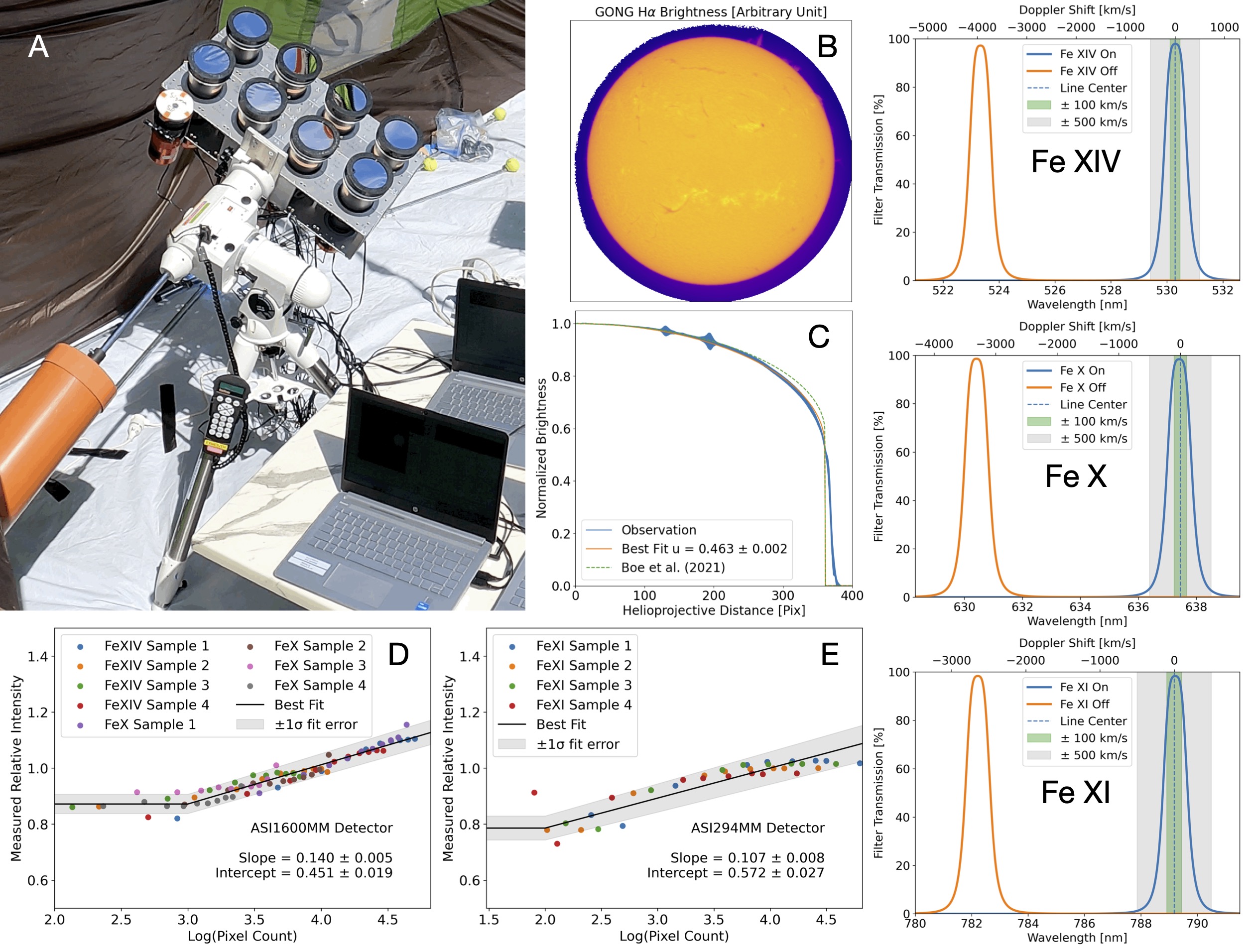}
\caption{Collection of information about the instrumentation and calibration of the Narrowband telescope systems. Panel A shows an image of the telescope setup in Exmouth. Panel B shows the GONG H-$\alpha$ brightness at the time of totality. Panel C shows the radial trace of the solar disk brightness observed for the \ion[Fe x] off-band ND observation used for absolutely calibrating the data. The right panels show the bandpasses for all the narrowband filters (on and off) for \ion[Fe xiv] (top), \ion[Fe x] (middle), and \ion[Fe xi] (bottom). Panels D and E show the detector linearity data for the ASI1600MM and ASI294MM cameras, respectively.}
\label{fig:Calib}
\end{figure*}

A picture of the observing setup in Exmouth is shown in panel A of Figure \ref{fig:Calib}. In order to maximize the number of telescopes per mount, we used a machined aluminum infrastructure to hold up to eight telescopes in a grid, with an additional slit-less spectrum camera mounted on the side (see on the lower left of the mount in the image). The Sky-Watcher EQ6 mount was pushed to the absolute weight limit with this setup, so we fashioned custom counter-weights constructed with PVC pipes filled with locally-sourced (and returned) rocks. For the mobile site on the Island, we used smaller Sky-Watcher EQ5 mounts that had only two telescopes per mount.
\par 

Each telescope was held within the aluminum array using 3D printed plastic holders that had adjustable tip/tilt screws to co-align the systems, as well as to adjust and lock the lens focus. Both the alignment and focus were optimized by observing the solar disk with all systems, using additional mylar filters to protect the instruments from the high intensity of the photosphere (equipped on the cameras in the image in Figure \ref{fig:Calib}). The tip/tilt was set by centering the Sun in the field-of-view (FOV) of all cameras simultaneously, while the focus was refined using software that measured the exact size of the solar disk as a function of the focus setting, which allowed nearly diffraction-limited focus for all systems. Each pair of cameras (i.e., each on/off-bandset for each line) was then operated by a single laptop. Using the same laptop enabled us to synchronize the times of exposure for each pair of on/off observations for the same line. In the event of light cloud cover, we planned to use the synchronization as a way to remove the effect of the cloud attenuation in each exposure, but that was not required since both observing sites had fantastic conditions during totality.

\par

After all the eclipse data had been co-aligned, we determined the correct orientation of the images (to have solar north upward) using H-$\alpha$ observations made by the GONG observatory at the Udaipur station in India \citep{Harvey1996, Hill2018}. We averaged all of the H-$\alpha$ data taken within a window of $\pm$ 15 minutes of the time of totality in Exmouth, resulting in the data shown in panel B of Figure \ref{fig:Calib}. Coincidentally, the Learmonth, Australia GONG station was less than forty kilometers away from our observing site in Exmouth, but since they experienced totality as well, the data from Udaipur were better for this alignment purpose. There was a rather large prominence in the northwest part of the solar disk, which was used to get a strong phase correlation match between the eclipse data (which also had that prominence) and the GONG data.

\par

\subsection{Narrowband Filters}
\label{sec:Filters}

As discussed in Section \ref{sec:Eclipse}, each emission line requires both an on-band observation centered on the line emission and a second off-band observation to quantify the nearby continuum. The exact filters used in this work are shown in the middle column of Figure \ref{fig:Calib} for \ion[Fe xiv] (top), \ion[Fe x] (middle), and \ion[Fe xi] (bottom). The widths of these bandpasses were increased to about 1-1.5 nm FWHM, in contrast to our earlier systems, which used 0.5 nm FWHM filters. The primary reason for the larger width is due to the expected increase in coronal line-widths as a function of height (e.g., \citealt{Esser1999, Boe2022}), so these wider filters are able to more accurately measure the brightness of the lines to large helioprojective distances. One additional advantage of the larger filter widths is that they do not require a heater assembly to ensure a precise temperature, while the older filters did due to their stronger sensitivity to temperature dependency in their transmission spectrum. The separation of several nanometers for the off-band filters was also essential, as prior observations with smaller separations of only 0.9-1.4 nm found Doppler shifted emission in the Off-band observations from a halo-CME \citep{Boe2020a}. The 7 nm separation is more than sufficient to ensure there will not be any possibility of Doppler shifted contamination (would require velocities $>$2000 km/s) while also remaining close enough in wavelength to prevent any F-corona related color effects on the background continuum.

\par

The reason we observed these three coronal lines in particular was due to their consistently being the brightest coronal lines in the visible and near-infrared (below 1000 nm), but also due to their response to electron temperature. The ionization equilibrium abundance of these three ions as a function of electron temperature (see Figure \ref{fig:Ionization}). These ions probe the majority of coronal plasma from 0.8 to 2.4 MK, which contains almost all regions in the corona other than the cores of high-temperature active regions and flares (see \citealt{Boe2023}). Thus, probing the emission of each of these lines enables an inference of the approximate coronal $T_e$ resolved for different structures in the corona.

\subsection{Calibration}
\label{sec:Calib}

To photometrically calibrate the data, we used observations of the solar disk taken after totality. In particular, we took observations with the \ion[Fe x] off-band system with additional neutral density (ND) filters that had a net attenuation of $2 \times 10^{-7}$, which comes from the custom-made $5 \times 10^{-6}$ filter combined with a reduction in aperture from 7.5 cm to 1.5 cm. We performed the solar disk ND observations between 05:23 and 05:27 UT (13:23 to 13:27 local time) when the Sun was at nearly the same altitude in the sky (within one degree) that it had been during totality -- during the eclipse, the Sun was still rising, and during the ND observations, it was setting. Therefore, the atmospheric attenuation experienced during the calibration will be nearly identical to the time of the eclipse, ensuring the best possible absolute calibration. We chose to perform the ND calibration with the \ion[Fe x] off-band telescope in particular due to the lower atmospheric scattering compared to the \ion[Fe xiv] continuum wavelength, and due to the fact that the longer wavelengths (near \ion[Fe xi]) typically have more issues with reflections introduced by the additional ND filter (e.g., Ghost images, see Section \ref{sec:Ghosts}).
\par

We then fit the limb-darkening profile of the solar disk in the ND data, as shown in panel C of Figure \ref{fig:Calib}. We found nearly the same limb-darkening coefficient we had with our older generation of telescopes in 2019 (see \citealt{Boe2021}). This fitting enabled a determination of the average solar disk brightness in units of pixel counts, as well as a determination of the size of the solar disk in the telescope. The solar disk data assisted in scaling the GONG H-$\alpha$ data prior to aligning the rotation, and determined the spatial scaling of helioprojective distance in the narrowband data. Once the \ion[Fe xiv] off-band had been calibrated into mean solar brightness units, we cross-calibrated the rest of the off-band observations (from both sites) using the observed brightness in a small region inside the largest streamer in the northeast (at about 1.1 \Rs \ and 30 degrees position angle). Since the K-corona dominates in such regions, we can assume that the actual brightness is nearly identical between all continuum channels -- whereas the F-corona dominates farther out and shows an increasing relative emission at longer wavelengths. Indeed, the relative emission between all continuum channels is rather consistent below about 1.5 \Rs, as shown in the bottom panels of Figure \ref{fig:Cont}. 

\par

Finally, we accounted for the sky background in the continuum data by subtracting the brightness in the very corner of the FOV (at about 7.5 \Rs). The line emission data do not need a background subtraction, as that is already accounted for in the off-band subtraction. However, for the continuum data, the sky brightness is not all of the signal even at that high helioprojective height. Thus, we first subtracted the background, then added back some brightness until the off-band \ion[Fe xiv] brightness matched the LASCO-C2 brightness in the northwest streamer (about 340 PA at 6 \Rs), given how closely the LASCO-C2 brightness matched our calibrated data below about 4 \Rs. Assuming the LASCO-C2 brightness is relatively accurate, we found that the coronal continuum brightness was only 20 $\%$ of the total brightness at about 7.5 \Rs. That is, the sky brightness was 80$\%$ of the signal at that large height.

\subsection{Detector Linearity}
\label{sec:Linearity}

The new detectors used in this work (as described in Section \ref{sec:Eclipse}) were all CMOS-style astrophotography detectors, designed for use by amateur astronomers. These detectors offer a rather high spatial resolution and FOV at an affordable price (given that several are needed for deployment at an eclipse), but they are not perfectly linear in their response. That is, the analog-to-digital converter in the camera is not completely uniform in its response depending on the number of counts in a given observation. To account for the non-linearity, we observed a constant light source from a calibration lamp with a large tungsten filament that was heated to a precise temperature (using a fixed current and voltage). We used an off-axis parabolic mirror in the lab to collimate light, simulating the same behavior as observing the Sun (i.e., a light source at infinity), and cycled over a range of exposure times to observe the exact same signal at different levels of saturation. 
\par
Next, we selected and averaged a number of specific regions containing hundreds of pixels over a perceptively uniform part of the filament structure in order to test how the signal varied as a function of total counts (via different exposure times). We did this experiment with an off-band telescope for all three ions to check for any wavelength dependence. The linearity data collected for the ASI1600MM camera is shown in panel D of Figure \ref{fig:Calib} for the \ion[Fe xiv] and \ion[Fe x] filters, and in panel E for the \ion[Fe xi] filter and ASI294MM camera. In those plots, we show the relative signal of each region normalized based on the exposure time (relative to the signal at $10^4$ counts) as a function of counts in log-space. Below about 1000 (100) counts, the ASI1600MM (ASI294MM) cameras do appear linear. However, as the counts increase, the relative intensity actually grows slightly. The ASI1600MM camera shows a more dramatic effect, with a change of about $30 \%$ over the dynamic range. We fit the data using a least-squares procedure, which is shown as the black line in the plots, with the 1-$\sigma$ uncertainty on the fit shown as a grey bar. We corrected all data (the flats, eclipse images, and ND observations) using the fitted functions shown in the plots prior to further processing, which corrects for the linearity while preserving the photometric calibration.

\par
\subsection{Ghost Images}
\label{sec:Ghosts}

As with any optical system that has multiple glass surfaces, there are inevitably reflections that occur in our telescopic systems. There are two primary sources of reflections that occurred, one from the narrowband filter and the other from the lens. Both of these reflections originate as a secondary backward reflection of the original reflection off the detector itself. These reflections can create ``ghost" images that are superimposed on top of the original image (often referred to as a lens flare). Thankfully, these reflected images are rather faint, typically less than $0.1-1 \%$ of the original signal, but even that amount can create substantial problems if the reflected image of the low corona lands in a part of the original image that is far from the Sun. 
\par

To eliminate the filter ghost image entirely, we tilted the filters by 1.5 degrees relative to the optical axis, which results in the ghost image landing outside the FOV of the detector. The lens ghost is more difficult to correct for, as it still lands inside the FOV of our data. To account for this ghost image, we self-subtracted the data with a scaling factor, physical offset, rotation, magnification, and Gaussian blur optimized to remove the ghost image from each telescope's data. The optimization was done painstakingly through a manual brute-force method (technique described in \citealt{Boe2020a}) until the signature of the ghost image was minimized. Attempts were made to automate the procedure, but the phase correlation signal was too weak to obtain as optimal a removal as could be achieved manually. To account for possible bias due to this process, an additional photometric error equal to the brightness of the ghost image was added. Nevertheless, this additional error does not have a significant impact on the results presented in this work.

\newpage 

\bibliographystyle{aasjournalv7}
\bibliography{2023Line.bib}

\end{document}